**Power Laws are Boltzmann Laws in Disguise**

by
Peter Richmond, Department of Physics, Trinity College Dublin 2, Ireland
and
Sorin Solomon, Racah Institute of Physics, Hebrew University of Jerusalem, Israel

**Abstract**
Using a model based on generalised Lotka Volterra dynamics together with some recent results for the solution of generalised Langevin equations, we show that the equilibrium solution for the probability distribution of wealth has two characteristic regimes. For large values of wealth it takes the form of a Pareto style power law. For small values of wealth, $w \leq w_m$ the distribution function tends sharply to zero with infinite slope. The origin of this law lies in the random multiplicative process built into the model. Whilst such results have been known since the time of Gibrat, the present framework allows for a stable power law in an arbitrary and irregular global dynamics, so long as the market is 'fair', i.e., there is no net advantage to any particular group or individual. We show for our model that the relative distribution of wealth follows a **time independent** distribution of this form even thought the total wealth may follow a more complicated dynamics and vary with time in an arbitrary manner.
In developing the theory, we draw parallels with conventional thermodynamics and derive for the system the associated laws of 'econodynamics' together with the associated econodynamic potentials. The power law that arises in the distribution function may then be identified with new additional logarithmic terms in the familiar Boltzmann distribution function for the system.
The distribution function of stock market returns for our model, it is argued, will follow the same qualitative laws and exhibit power law behaviour.

**Background**
Empirical studies of data in many areas reveals statistical distributions that follow Pareto power laws as opposed to Gaussian distributions. For example, within countries, the number of cities, $n$, with population between $p$ and $p+dp$ is given by

$$n(p)dp \sim dp / p^{1+a} \quad (1.1)$$

The parameter $a$ in this case is close to unity[1]. The distribution of words or letters, ranked according to useage also follows a power law[2]. Many more examples are to be found in biology and economics. The rank order of businesses by turnover and universities by research income or patents all follow power laws [3].

In quantitative finance, many empirical studies reveal that probability distribution functions for stock fluctuations have 'fat tails' that follow power laws prior to truncation. Mandelbrot was the first to observe this, using daily prices for cotton in 1963 [4]. Yet since 1900 and the work of Bachelier [5], theories based on Brownian motion, that grossly underestimate large amplitude fluctuations have been widely used in attempts to model stock fluctuations. Their continued use seems to emanate from the development of the Ito stochastic calculus that facilitated the development of the famous Black-Scholes equation used to model financial derivatives. Conversely the fitting of stock fluctuations by Levy distributions overestimates the extent of the fluctuations with its prediction of infinite volatility.

Agent models[6] have begun to be used by physicists and numerical studies suggest that these can model both the power law tails and the 'clustered volatility'. Analytic approaches have so far been less successful. However, in a series of papers, Solomon and co-workers[7][8] have suggested that a very wide class of simple microscopic representation models based on generalised Lotka-Volterra (GLV) dynamics can account for the generic properties of financial markets. The main conceptual ingredient responsible for the emergence of power laws in these models is the auto-catalytic character of the capital dynamics, i.e., the time variation of the capital (returns) is a random quantity proportional to the capital itself. This implies mathematically that the dynamics of such systems is governed by random multiplicative processes. These have been known since the time of Gilbrat[9], however the present framework allows for a stable power law in an arbitrary global dynamics as long as the market is 'fair', i.e., there is no net advantage to any agent or microscopic group.

Recently, Richmond[10] has studied a class of generalised Langevin equations for which the associated probability distributions functions can be obtained analytically. The result illustrates explicitly how the power law emerges in a natural manner and may be reinterpreted as a logarithmic element of a 'thermodynamic' potential. In this paper we extend this approach and apply it to the microscopic agent models previously studied by Solomon et al. The probability distribution functions can be derived analytically and validate the numerical results. In addition we derive new results that show for an infinite number of interacting agents that the distribution of relative wealth of the agents is independent of time even though the total wealth of the system may be varying in an unknown and arbitrary manner.

Section 2 summarises the GLV formalism of Solomon. The results of Richmond are derived using a simpler approach and by analogy with thermodynamics, some new econodynamic functions are defined for this model in section 3. In section 4 we derive new results relating to the relative wealth distribution function. These show that under a wide range of conditions the wealth distribution function is independent of time. This surprising result has deep implications for those concerned with setting policy relating to economic and social systems and some of the aspects are discussed.

**2. General Framework**
In differential form, the GLV description can be expressed as follows.

$$\frac{dw_i}{dt} = [\mathbf{l}(t) - 1]w_i + aw - cww_i \qquad (1.2)$$

where

$$w = \frac{1}{N}\sum_j w_j \qquad (1.3)$$

Each agent has wealth, $w_i$. We assume that the stochastic term follows the usual Gaussian behaviour

$$\langle l(t) \rangle = \bar{I}$$
$$\langle (l(t) - \bar{I})(l(t') - \bar{I}) \rangle = 2D\delta(t - t') \tag{1.4}$$

The second term on the LHS of (1.2) which is proportional to the average wealth of the agents prevents, as we shall show, the individual wealth falling below a certain minimum fraction of average. The exact mechanism by which this happens, subsidies, minimal insurance or wage, elimination of the weak and their substitution by the more fit is not, at this level of description, important. The third term, which controls the overall growth of the wealth in the system, represents external limiting factors (finite amount of resources and money in the economy, technological inventions, wars, disasters etc.) as well as internal market effects (competition between investors, adverse influence of bids on prices such as when large investors sell assets to realize their value and prices plus returns fall as a result.

We may rewrite the GLV as follows:
$$\frac{dw_i}{dt} = [(l(t) - \bar{I})]w_i + m(w)w_i + aw \tag{1.5}$$

We have introduced
$$m(w) = -cw - 1 + \bar{I} \tag{1.6}$$

## 3. Econodynamics

At this point we digress to give a shorter derivation of the result derived elsewhere for the probability distribution function associated with the Langevin equation, (1.5). This not only allows us to quickly derive the 'equilibrium' distribution function, but also identify the 'econodynamic' potential functions by analogy with conventional thermodynamics..

Consider the generalised Langevin equation

$$\dot{x} = F(x) + G(x)h(t) \tag{1.7}$$

where
$$\langle h(t) \rangle = 0 \tag{1.8}$$

and
$$\langle h(t)h(t') \rangle = 2D\delta(t - t') \tag{1.9}$$

Introduce H and V where
$$\frac{dH}{dt} = \frac{1}{G}\frac{dx}{dt} \text{ and } -\frac{\partial V}{\partial H} = \frac{F}{G} \text{ or } -\frac{\partial V}{\partial x} = \frac{F}{G^2} \tag{1.10}$$

Equation (1.7) thus may be written as
$$\frac{dH}{dt} = -\frac{\partial V}{\partial H} + h \tag{1.11}$$

This has the standard form of a random walk and the equilibrium solution for the probability distribution B(H) is given by
$$B(H)dH = \frac{1}{\mathbb{N}}\exp[-V(H)]dH \tag{1.12}$$

In terms of the original variable, $x$, we have

$$p(x)dx = B(H)dH$$
$$= \frac{1}{\mathbb{N}} \exp[\frac{1}{D} \int \frac{F(H')}{G(H')} dH'] dH \qquad (1.13)$$
$$= \frac{1}{\mathbb{N}} \exp[\frac{1}{D} \int \frac{F}{G^2} dx'] \frac{dx}{G}$$

This yields

$$p(x) = \frac{1}{\mathbb{N}} \exp[-\{\frac{V(x)}{D} + \ln G\}] \qquad (1.14)$$

Now introduce the 'econodynamic' entropy

$$S = -\langle \ln p(x) \rangle \qquad (1.15)$$

We then obtain

$$F = U - DS \qquad (1.16)$$

The effective 'internal' energy is

$$U(x) = V + D \ln G \qquad (1.17)$$

The 'Hemholtz' Free energy is

$$F = -D \ln \mathbb{N} \qquad (1.18)$$

$\mathbb{N}$ is the partition function for our model. We see in (1.7) that the effect of the modulation of the fluctuations by the function G is to introduce an additional contribution to the internal potential. This term, as we shall see, gives rise to the power laws in the distribution function.

Choosing the functions G and F to be of the form $G(x) = x$ and $F(x) = k - px$, it follows that

$$p(x) = \frac{1}{\mathbb{N}} \frac{\exp[\frac{1}{D} \int \frac{(k-px)}{x^2} dx]}{x} \qquad (1.19)$$
$$= \frac{1}{\mathbb{N}} \frac{\exp[-k/(Dx)]}{x^{(1+p/D)}}$$

The 'free' energy is

$$U(x) = k/x + (p+D) \ln x \qquad (1.20)$$

Both p and U are shown schematically in the figures below. The distribution function clearly decays to zero as a power law with index 1+p/D for large values of x and, interestingly, goes to zero, as x tends to zero with infinite slope.

Insert Figure 1 here

Specifically, for the GLV model, $G = w_i$ and $F = k - pw_i$ where $k = aw$ and $p = cw - \bar{I}$. Thus

$$p(w_i) = \frac{1}{\mathbb{N}} \frac{\exp[-\frac{aw}{Dw_i}]}{w_i^{\{1+\frac{(cw-\bar{I})}{D}\}}} \tag{1.21}$$

Stochastic Lotka-Volterra systems of competing auto-catalytic agents lead generically to truncated Pareto power wealth distribution, truncated levy distribution of market returns, clustered volatility, booms and crashes[11].

It was shown in reference 11 that in the GLV model used here $w$ may be replaced by its mean value, $\bar{w}$. Furthermore, since $\bar{w}_i = \bar{w}$ we have

$$\bar{w}_i = \bar{w} = \frac{1}{\mathbb{N}} \int_0^\infty \frac{e^{-\frac{a\bar{w}}{Dx}}}{x^{\frac{(c\bar{w}-\bar{I})}{D}}} dx \tag{1.22}$$

The partition function

$$\mathbb{N} = \int_0^\infty \frac{e^{-\frac{a\bar{w}}{Dx}}}{x^{1+\frac{(c\bar{w}-\bar{I})}{D}}} dx \tag{1.23}$$

We now use the result

$$\int_0^\infty \frac{\exp(-k_2/x)}{x^{k_2}} dx \bigg/ \int_0^\infty \frac{\exp(-k_2/x)}{x^{1+k_2}} dx = k_2/(k_1-1) \tag{1.24}$$

and obtain after a little algebra

$$c\bar{w} - \bar{I} = a + D \tag{1.25}$$

Substituting this result into equation (1.21) gives

$$p(w_i) = \frac{1}{\mathbb{N}} \frac{\exp[-\frac{a\bar{w}}{Dw_i}]}{w_i^{2+\frac{a}{D}}} \tag{1.26}$$

Note this result is independent of both c and $\bar{I}$! The distribution function tends to zero as $w_i$ tends to both zero and infinity. It has a single maximum value when $dp/dx_i = 0$. The value $w_m$ for which $p(w)$ is a maximum may be readily calculated and this yields:

$$Q = \frac{w_m}{\bar{w}} = \frac{1}{(2D/a+1)} \tag{1.27}$$

## 4. Time Dependent Lotka Volterra models

Let us now consider a more general case where the system does not necessarily have a fixed value of total wealth, $\bar{w}$. Consider again the equation (1.5) and let us assume that the function $m$ is of a more general character that may be time-dependant, i.e.,

$$\frac{dw_i}{dt} = [(\mathbf{l}(t) - \bar{\mathbf{l}})]w_i + m(\bar{w},t)w_i + aw \qquad (1.28)$$

The term $m(\bar{w},t)$ represents the general state of the economy. Time periods where $m(\bar{w},t)$ is large and positive correspond to boom periods in where the wealth of individuals is, on average increasing, while a period where $m(\bar{w},t)$ is negative corresponds to a slow down of the economy when typically the investments lead to negative or small returns. The main message of the present paper is that, as long as the term $m(\bar{w},t)$ is the same for all the equations for arbitrary $i$, the Pareto power law holds and its exponent is independent of the nature of $m(\bar{w},t)$.

Consider now the relative wealth $x_i = w_i / \bar{w}$. Differentiating yields

$$\frac{dx_i}{dt} = \frac{1}{\bar{w}}\frac{dw_i}{dt} - \frac{w_i}{\bar{w}^2}\frac{dw}{dt} \qquad (1.29)$$

From (1.28) and (1.29) we obtain

$$\frac{dx_i}{dt} = [(\mathbf{l}(t) - \bar{\mathbf{l}})]x_i + a + \left(m(\bar{w},t) - \frac{d\bar{w}/dt}{\bar{w}}\right)x_i \qquad (1.30)$$

It now follows invoking the result (1.13) that the probability distribution function for the relative wealth, $p(x_i)$ is

$$p(x_i) = \frac{\exp\{-a/(Dx_i)\}}{\mathbb{N}x_i^{[1-\left(m(w,t) - \frac{d\ln\bar{w}}{dt}\right)]}} \qquad (1.31)$$

Imposing the identity $\langle x_i \rangle = 1$ gives

$$\frac{1}{\bar{w}}\frac{d\bar{w}}{dt} = a + D + m(\bar{w},t) \qquad (1.32)$$

If solutions exist to the equation

$$a + D + m(\bar{w},t) = 0 \qquad (1.33)$$

then $\bar{w}$ may fall into one of these solutions. If not, $\bar{w}$ may have an eventful history running forever to zero, infinity or wandering irregularly between them. However, it is interesting to note that, using the solution (1.33), the probability distribution function may be re-expressed in a form that is independent of the function, $m$, i.e., :

$$p(x_i) = \frac{\exp[-\frac{a}{Dx_i}]}{\mathbb{N}x_i^{(1+\mathbf{a})}} \qquad (1.34)$$

The exponent

$$\mathbf{a} = 1 + a/D \qquad (1.35)$$

The important feature of this result is that it is totally time independent regardless of the form of the function $m$. The total wealth, $\bar{w}$, may thus vary in a complicated manner according to the dynamical equations (1.32), however, the distribution of relative wealth remains fixed forever!

From the shape of the probability distribution function it should be clear that whilst the decay to zero for large values of the argument is slow, the decay to zero for small values of the argument is very fast. As we have already remarked, the slope is infinite as the argument goes to zero. This means that to all intents and purposes, the value of the argument at the maximum point is the effective minimum value for the relative wealth. In terms of this minimum value:

$$x_m = (a-1)/(a+1) \text{ i.e. } a = (1+x_m)/(1-x_m) \tag{1.36}$$

The key result we obtain is, therefore, that the lower relative wealth bound totally governs the overall distribution function [12]. The dynamics by which this distribution arises are of course complex and depend on the interactions in the system.

In societies mechanisms such as collective bargaining, strikes, speculation and investment insure the distribution has the power law form. It is these kind of economic interactions that ensure the power law is modified so as to maintain the appropriate minimum level of income.

One might consider what this minimum wealth might be. A key criterion for negotiation is the need for survival and support immediate dependents. Assuming a family with one wage earner and 3 dependants suggests that $x_m \sim 1/4$, a value not too far away from actual values in many societies. The lowest income is around a quarter of the average value. From equation (1.36) it follows that $a = 5/3 \sim 1.67$. The low birth rate in some of today's societies might suggest higher values for $x_m$ with associated lower values for $a$ leading to greater equality and stability. If the fluctuations in the economy are large, the social subsidies, as measured by $a$, need to be larger to ensure both $x_m$ and $a$ remain constant. For example, energetic stock markets combined with stagnant social security or pensions may lead to a decrease in $a$. The subsequent increase in inequality can give rise to larger market and social fluctuations.

Clearly our approach is of a simple character. Non the less we feel that it offers the beginnings of a more innovative approach to economic dynamics and might even at this stage offer some new ideas for policy makers. It suggests also that political leaders might focus more on macroeconomic as opposed to microeconomic factors in setting medium term policy goals together with ensuring, via appropriate regulation, that societal interactions create 'liquidity'. Intervention and manipulation of microeconomic factors may offer, at best, short-term solutions to economic problems and at worst may accentuate the avalanche style shocks within the economy as the system seeks to reach a 'steady or natural state' consistent with the macro parameters.

**Stock returns**
Finally we use our approach to examine the relationship between the relative wealth of our agents and the total wealth returns. These would govern the fluctuations of say the stock market price. The total return, $R$, is given by

$$R(t,\tau) = \ln[w(t)/w(t-\tau)]$$
$$= \int_{t-\tau}^{t} ds \frac{1}{w} \frac{dw}{ds} \qquad (1.37)$$
$$= \int_{t-\tau}^{t} ds \frac{1}{w} \sum_i \frac{dw_i}{ds}$$

Substituting for $\dot{w}_i$ using equation (1.28) gives

$$R(t,\tau) = \int_{t-\tau}^{t} ds \left( \frac{1}{N} \sum_i \eta x_i + m + a \right) \qquad (1.38)$$

The second and third terms on the RHS of equation (1.38) represent a deterministic trend that depends on the detailed form of the function, $m(\bar{w},t)$. However, the first term is of a stochastic nature. The summand, $\eta x_i$, is distributed in principle by a power law with exponent $1+\alpha$. Consequently, the sum is distributed by a Levy distribution with index $\alpha$. Therefore the absolute value of the sum is of order $N^{\frac{1}{\alpha}}$. The absolute value of $R(t,1)$ is then of order $N^{\left(\frac{1}{\alpha}-1\right)}$. This fits what we know about the extreme cases. For $\alpha = 2$, one obtains the Gaussian case, i.e. $R \sim N^{-1/2}$. For $\alpha = 1$, one obtains $R \sim O(1)$ and the fluctuations remain macroscopic even in the limit $N \to \infty$.

**References**
[1] See for example: A Blank and S Solomon Physica A 287 (2000) 279-288
[2] U G Yule Phil Trans B 213 (1924) 21
[3] M H R Stanley, L A N Amaral, S V Buldyrev, S Leschhorn, P Maass, M A Salinger and H E Stanley Nature **379** 1996 804-806
[4] B B Mandelbrot Journal of Business **38** (1963) 394-419
[5] L Bachelier, Theorie de la Speculation Paril 1900 translated and reproduced in The Random Character of Stock Market Prices, P H Cootner (Ed) MIT Press 1964
[6] For a survey, see M Levy, H Levy and S Solomon Microscopic Simulation of Financial markets, Academic Press, New York, 2000
[7] S Solomon and M Levy Int Journal of Mod Phys C 1996 7(5)
[8] O Biham, O Malcai, M Levy and S Solomon Phys Rev **E 58** (1998) 1352
[9] R Gilbrat, Les Inegalities Economiques, Sirey, Paris 1931
[10] P Richmond, Eur J Phys, (In press and in proceedings of APFA2 Liege 2000)
[11] S Solomon In 'Decision Technologies for Computational Finance' 73-86, Eds. A-P. N. Refenes, A.N. Burgess, J.E. Moody, (Kluwer Academic Publishers, Netherlands 1998)
[12] P W Anderson In 'The Economy as an Evolving Complex System II (Redwood City, Calif, Addison Wesley 1995) Ed W Brian Arthur, Steven N Durlauf and David A Lane

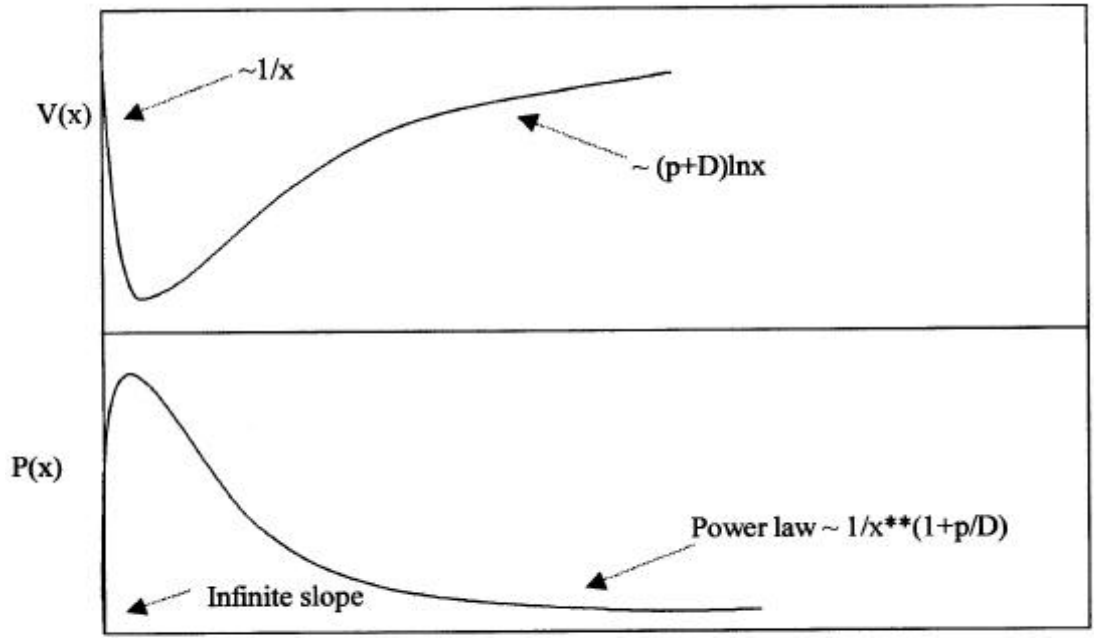

Figure 1 Schematic of the potential function, V and the distribution function, P, referred to above